# An HST Survey of the Disk Cluster Population of M31. II. ACS Pointings[1]

O. K. KRIENKE
Seattle Pacific University, Seattle, WA  okk@spu.edu

AND

P. W. HODGE
University of Washington, Seattle, WA  hodge@astro.washintgon.edu

ABSTRACT. This paper reports on a survey of star clusters in M31 based on archival images from the *Hubble Space Telescope*. Paper I (Krienke and Hodge 2007) reported results from images obtained with the Wide Field/Planetary Camera 2 (WFPC2) and this paper reports results from the Advanced Camera for Surveys (ACS). The ACS survey has yielded a total of 339 star clusters, 52 of which, mostly globular clusters, were found to have been cataloged previously. As for the previous survey, the luminosity function of the clusters drops steeply for absolute magnitudes fainter than $M(V) = -3$; the implied cluster mass function has a turnover for masses less than a few hundred solar masses. The color- integrated magnitude diagram of clusters shows three significant features: (1) a group of very red, luminous objects: the globular clusters, (2) a wide range in color for the fainter clusters, representing a considerable range in age and reddening, and (3) a maximum density of clusters centered approximately at $V = 21$, $B - V = 0.30$, $V - I = 0.50$, where there are intermediate-age, intermediate-mass clusters with ages close to 500 million years and masses of about 2000 solar masses. We give a brief qualitative interpretation of the distribution of clusters in the CMDs in terms of their formation and destruction rates. (key words: star clusters and associations; galaxies)

## 1. INTRODUCTION

This is the second paper in a series of two that reports the results of the identification and photometry of star clusters in the disk of the galaxy M31 (NGC 224). Paper I (Krienke and Hodge 2007) presented details of a set of 343 clusters that were detected and measured on *Hubble Space Telescope* Wide Field Planetary Camera 2 images. The present paper reports on *HST* Advanced Camera for Surveys images.

A history of the searches for disk clusters in M31 was summarized in Paper I. Except for bright globular clusters, only a few young clusters in M31's disk had been studied in ground-based surveys. The crowding and the faint magnitudes required the characteristics of the *Hubble Space Telescope* to make reliable identifications and measurements. Several important recent papers, listed in Paper I, have dealt with unusually luminous young clusters ("blue globular clusters") based on HST images. However, previous work on traditional "open clusters" in M31 has been more limited. The most important recent paper published before our Paper I is that of Barmby and Huchra (2001), who identified 20 probable open clusters and showed HST images of them.

We note here that, as in Paper I, we adopt the following parameters for M31: distance: 725 kpc, position angle of major axis: 37 degrees, inclination angle: 12.5 degrees to the line-of-sight, mean foreground reddening: $E(B-V) = 0.20$, mean internal reddening: $E(B-V) = 1.0$. The internal reddening in the optical in M31 is highly variable because of the relatively high dust content and the small inclination angle. Our survey of galaxies seen through the M31 disk on the ACS images, which we hope to publish soon, indicates that the optical depth in some of our areas is highly-structured.

## 2. OBSERVATIONAL MATERIAL

For this survey we chose 27 pointings from the list (as of 1 January 2007) of ACS exposures included in the *HST* archives. We made the following requirements: (1) there should be at least two colors from the various HST filters that provide colors convertible to B, V or I, (2) the location should be in the main disk of M31, though we chose two outer pointings for comparison, (3) the exposures should be at least 400 seconds (with the exception of dataset j92ga7vlq, for which the F555W exposure was only 151 seconds) and (4) the star density should be such that clusters would not be swamped by field stars, as is the case in the inner bulge and the nucleus region. When there were several pointings in the same region, we chose representative selections. The total area scanned was 278.37 $(arcmin)^2$. Figure 1 shows the distribution of the chosen pointings on the disk of M31 and Table 1 lists the pointings, their positions, filters and exposures. Because of the fact that we were limited to archival data, with filters and exposures dictated by other considerations, the ACS sample is somewhat less uniform in characteristics than the WFPC2 sample.

We located and measured a total of 339 star clusters. The search techniques and measurement procedures for this program were identical to those of Paper I. Star clusters were identified independently by each of the authors using the criteria described in Paper I. Candidates were accepted only if the images were unambiguous in both colors. As in the first survey (Paper I), there was an approximately equal number of possible, but doubtful, clusters noted. They are not tabulated or discussed here, as their true nature will require further observational tests.

Positions, integrated magnitudes and colors were obtained for each cluster, as described in Paper I. The largest uncertainty in the photometry is caused by the variable background of the M31 disk, which contributes significantly to the measurements for faint clusters. As in Paper I, for each cluster we measured the surface brightness and color of a large number of nearby background areas the same size as the adopted cluster size, allowing a determination of the background and its uncertainty. The colors and magnitudes were converted from the observed filters and exposures to standard B, V, and I values by using the method given by Siriani et al. (2002). Table 2 lists the mean measurement errors as a function of cluster integrated magnitude and color.

Comparing our data with previous catalogs, primarily the Revised Bologna Catalog of Clusters in M31 (hereafter, RBC) (Galleti et al. 2004) and the more recent list of new discoveries by Kim et al. (2007), we find that there are 52 clusters in common, most of them classified as globular clusters. Nine of them are rediscoveries of open clusters first observed by Hodge (1979). Table 3 lists the previously cataloged clusters and provides our photometry for them. Average differences in the sense of our values minus the RBC values are as follows: delta V = 0.25 +/- 0.08, delta B = 0.10 +/- 0.07, delta I = 0.26 +/- 0.12. We point out that these values are all positive, which is likely the result our use of small apertures to minimize the effects of the background on cluster colors. Some of the differences may also be the result of the fact that some of the RBC measurements come from older photographic values.

Table 4, which in its entirety is available in the electronic edition of the Publications, lists all of the new clusters, their positions, their integrated magnitudes and their colors. We have continued the numbering of clusters from Paper I; thus, the first of the new clusters is named KHM31-284.

Figure 2 shows a sampling of clusters, illustrating the range in size, structure, and brightness of the sample.

### 3. CLUSTER PROPERTIES

### 3.1. The luminosity Function

The observed luminosity function for the new clusters listed in both this paper and Paper I is shown in Figure 3. There is a dearth of bright globular clusters, as they had already been cataloged. The distribution shows a turn-over at an absolute magnitude of approximately M(V) = -3.0. It should be noted that the derived luminosity function may be biased by the fact that there is a range of exposure times in our sample. However, our tests of cluster detectability for images with different exposure times do not show a correlation. Apparently the number of clusters detected is more influenced by intrinsic and extrinsic effects (formation rates, extinction, etc.) than by exposure time.

### 3.2 Color-Magnitude Diagrams

In Figure 4 we plot the observed integrated magnitudes and colors for the two ACS samples, those with B and V colors and those with V and I. These diagrams have very similar distributions of points to those found for the sample of clusters in Paper I. There are three interesting features of all of these color-magnitude diagrams (CMDs):

1. the colors show a wide range of values, suggesting the presence of clusters of many ages, though the large uncertainties of colors for the faintest clusters weakens this conclusion for them.
2. there are a few very luminous, red clusters that stand out from the rest. These are the traditional globular clusters.
3. the fainter ("open") clusters show a distribution with a maximum density in the middle part of the diagram, where the typical intermediate-age, intermediate-mass clusters with ages close to 500 million years and masses of about 2000 solar masses (Girardi 2006). We interpret this concentration to be the result of an approximately constant rate of cluster formation plus reddening of clusters with age. This would predict few very blue (young) clusters and few very red ones (because of aging or disruption). We have discussed both of these effects in Paper I and the new results are in agreement with those presented there.

We note that our CMD for clusters shows a very different distribution from that of a CMD for a typical sampling of stars in the M31 disk (e.g., Magnier et al. 1997 and Williams 2003), which shows a more evenly-spread distribution of points, without the bunching of points at intermediate colors exhibited by the clusters. As explained above, this is the result of the fact that the clusters concentrate stars of a limited range of ages, because of dynamical evolution, so that the age distribution of clusters is skewed compared to that of field stars.

These CMDs plot the observed colors and magnitudes, uncorrected for reddening. The reddening correction for these clusters cannot be measured with existing data, as we have only two colors available for all but three of the pointings, and, in any case, the large variations in the background would make the uncertainties in colors too large for precise reddening determinations.

## 4. THE SPATIAL DISTRIBUTION

Figure 5 shows the distribution of the cluster density (corrected to face-on) in the disk as a function of distance from the center. We have combined the data with that of Paper I and have removed the globular clusters from this diagram, assuming them to belong to a different dynamical system. The diagram shows a relative paucity of clusters within 5 kpc (although the number of fields sampled there is small), a maximum density in the range 8 to 15 kpc and a gradual decrease beyond 15 kpc. Also, as for the sample of Paper I, the cluster density is correlated with the distance from the nearest star-forming region.

## 5. CONCLUSION

These two surveys of M31, using two cameras of the *Hubble Space Telescope*, have shown that the disk of M31 contains a large number of star clusters with a wide range of observable properties (luminosity, size, color) and a wide range of implied characteristics (mass, age, dynamical history). The total area scanned for clusters in the two papers is 477.4 (arcmin)$^2$, which is 4.9% of the area of the disk (depending on how the disk's boundaries are defined). The implied total number of disk clusters is on the order of $10^4$, indicating that the clusters offer a rich vein of material for the study of the history and mechanisms of star and cluster formation and destruction. To mine this vein will require accurate measurements of individual clusters' reddenings and extinction. We have begun a program that examines several methods of accomplishing this with existing data.

We are indebted to the American Astronomical Society's Small Grant Program for publication funds and to NASA and the Space Telescope Science Institute for making the archival images available.

Figure Captions

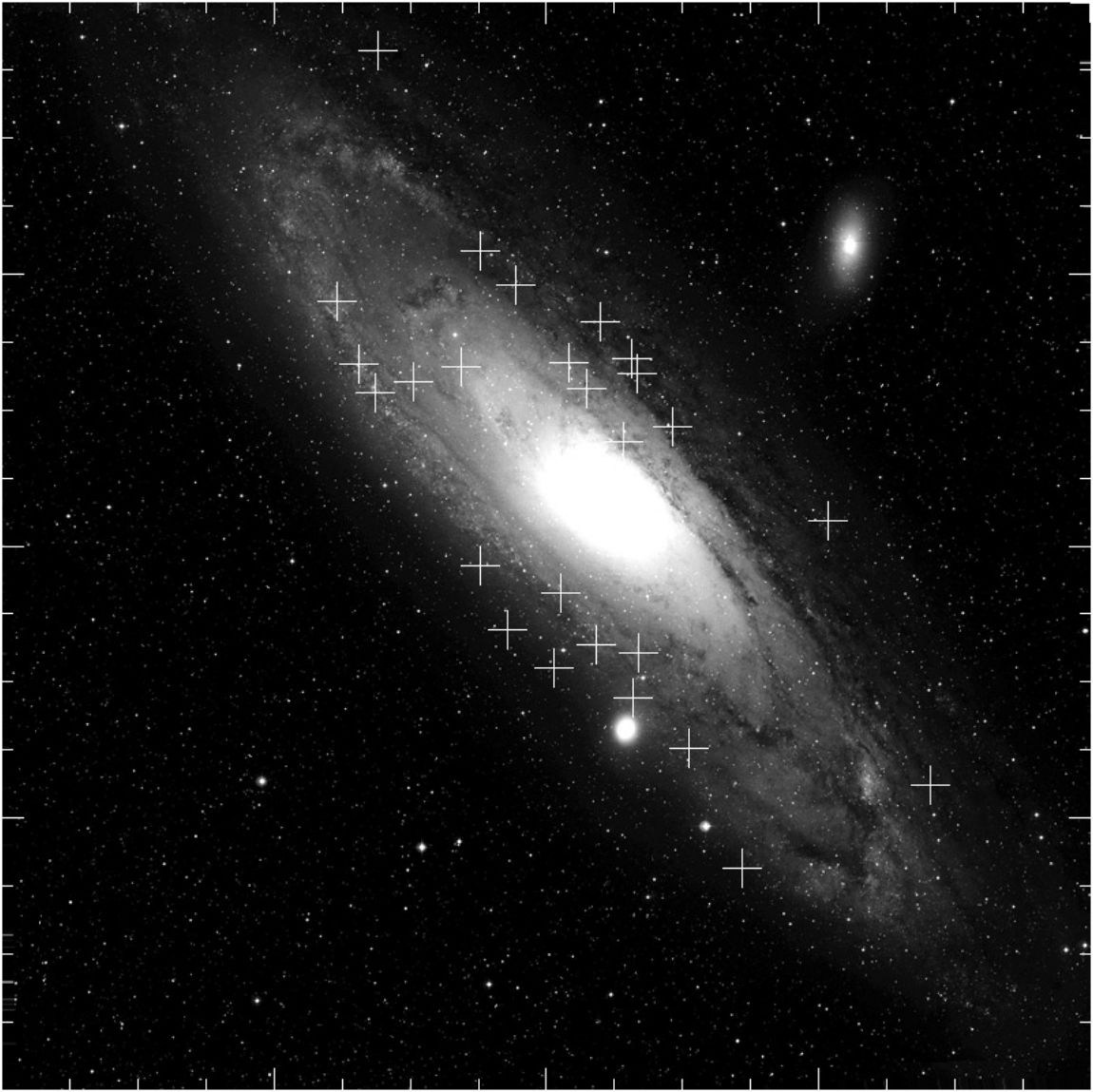

FIG. 1. – The locations of the *HST* ACS pointings that were used to search for star clusters. The background photograph is as described in Paper I. Two outer fields are beyond the edges of the figure (see Table 1). The width of the image is 85 minutes of arc. Note that the symbols are larger than the size of the ACS fields.

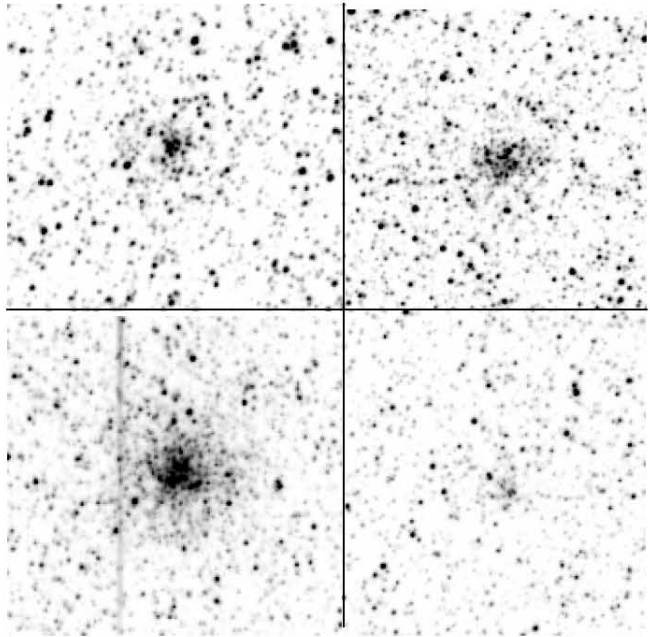

FIG. 2. – Four sample disk clusters. The upper left image is of KHM31-538 with an integrated magnitude of V = 20.97. The upper right cluster is KHM31-399, with V = 21.08. The lower left cluster, one of the brightest, is KHM31-409, with V = 18.87. The vertical streak is caused by a very luminous star directly above the field. The lower right panel shows one of the faintest clusters, KHM31-395, with V = 22.64. Each image is 17.5 arc seconds wide.

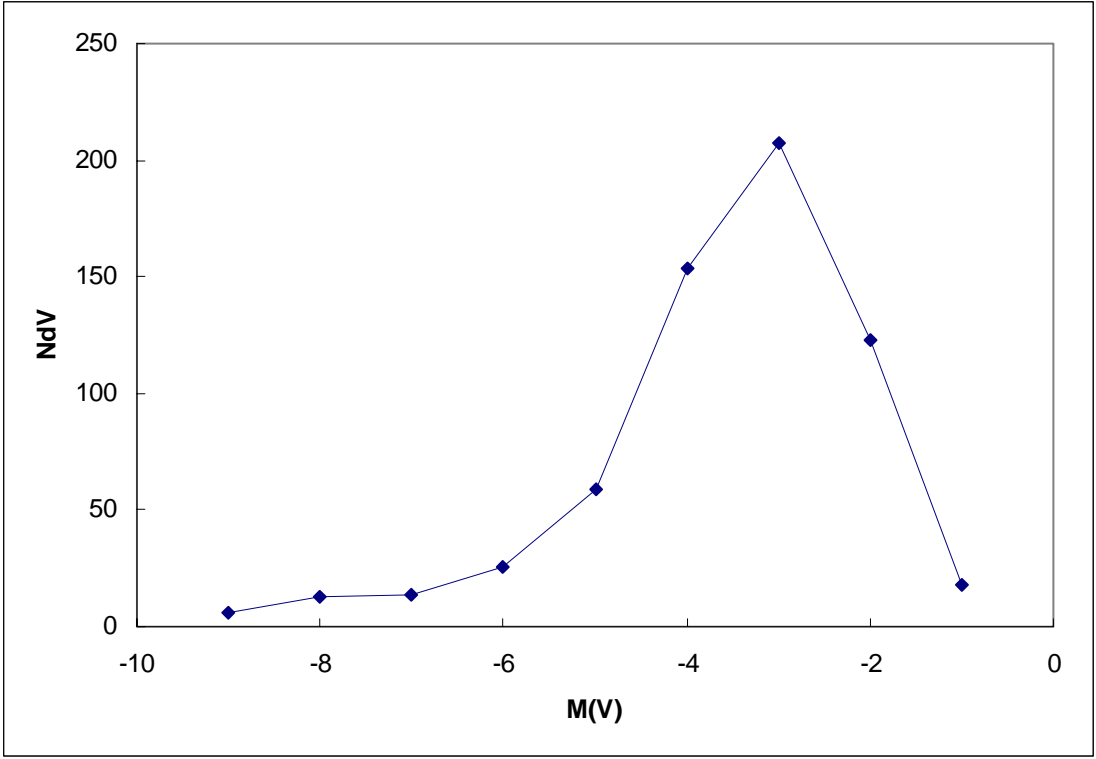

FIG. 3. – Integrated luminosity function of the new clusters measured in Paper I and this paper. The dots are connected by straight lines for clarity.

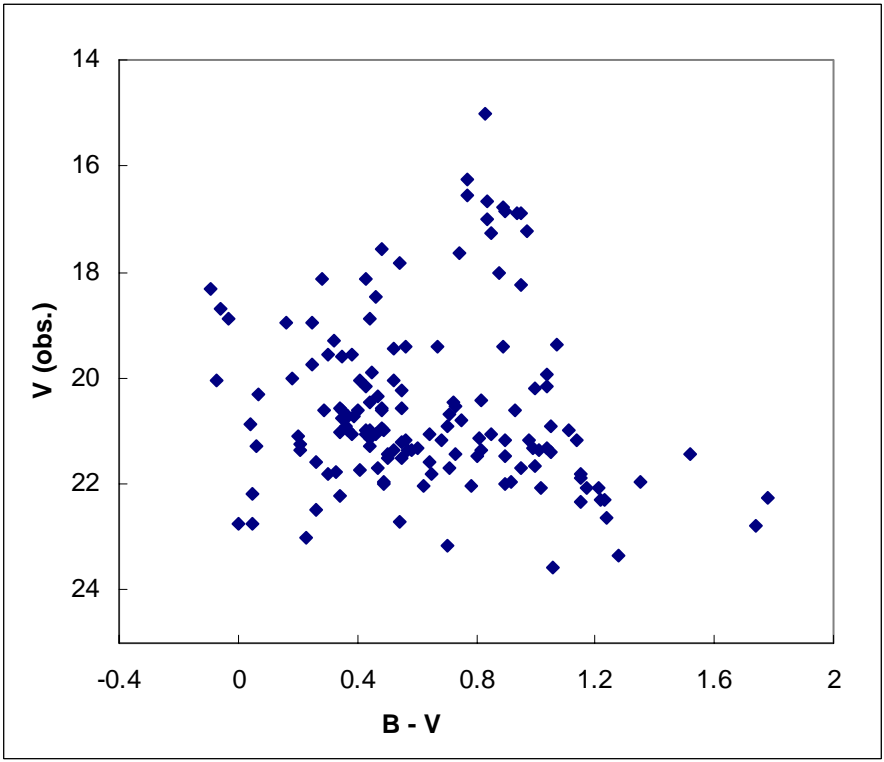

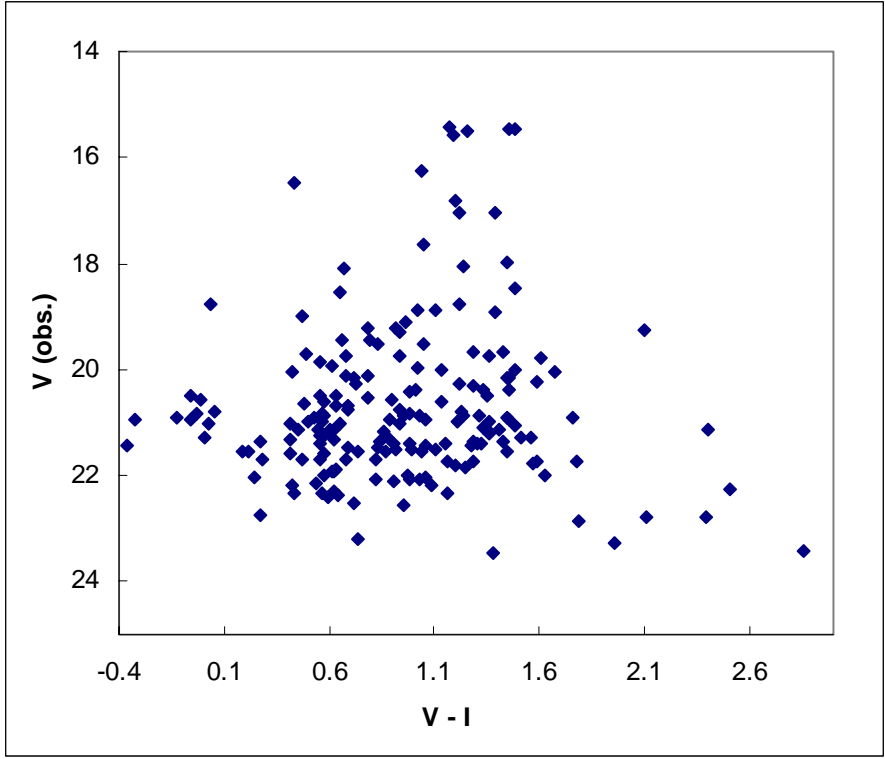

FIG. 4. – Integrated color-magnitude diagrams for all clusters detected in the ACS survey (Tables 2 and 3). The globular clusters occupy the upper right-hand area of the arrays.

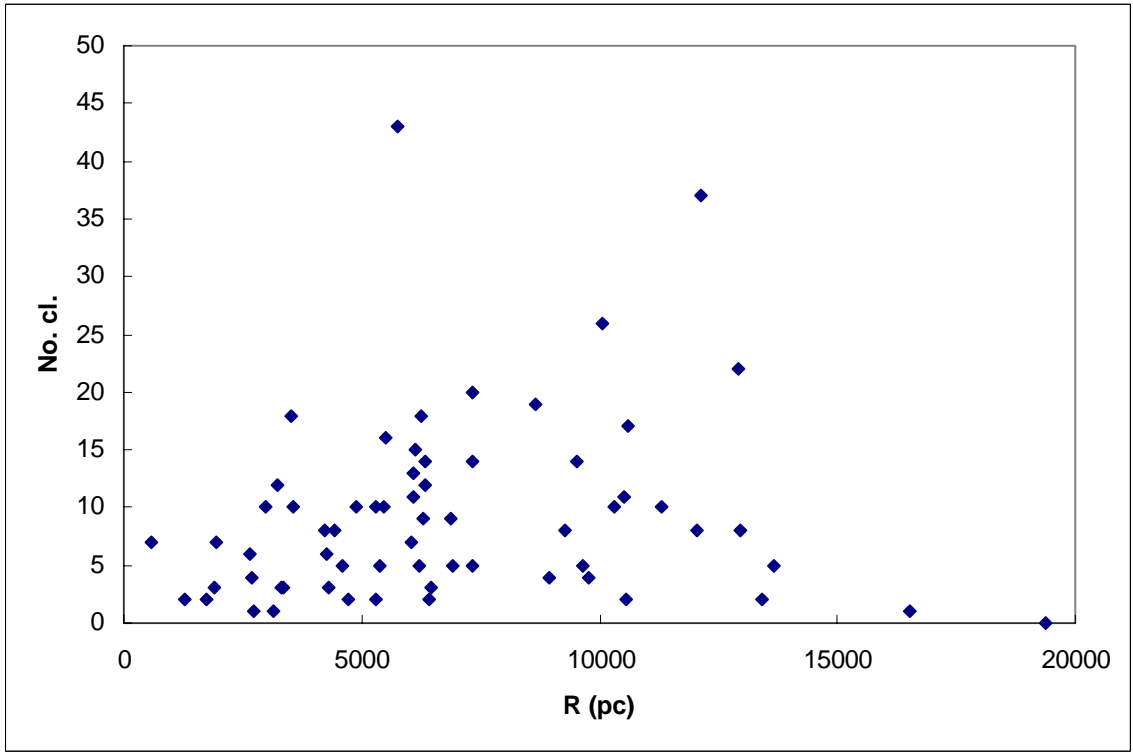

FIG. 5. – The radial distribution of clusters in the survey. We plot the number of clusters per field as a function of distance from the M31 nucleus, assuming all to be in the plane.

Tables

TABLE 1
*HST* ACS POINTINGS SEARCHED

| Dataset | RA | Dec | Ncom-bine | Filter 1 | Filter 2 | Exp 1 | Exp 2 |
|---|---|---|---|---|---|---|---|
| j8mg01071 | 9.884423 | 40.78762 | 3 | F555W | F814W | 870 | 825 |
| j96q07010 | 10.12638 | 41.25184 | 4 | F606W | F435W | 3250 | 7260 |
| j8mg04071 | 10.32099 | 40.64206 | 3 | F555W | F435W | 825 | 1200 |
| J96Q06010 | 10.44707 | 40.85222 | 3 | F606W | F435W | 2110 | 2672 |
| j92gb1d3q | 10.49058 | 41.41872 | 1 | F555W | F814W | 413 | 502 |
| j8z004010 | 10.56584 | 41.02121 | 3 | F606W | F814W | 2370 | 2370 |
| j92gb2hyq | 10.5765 | 41.51019 | 1 | F555W | F814W | 413 | 502 |
| j92gb9brq | 10.57663 | 40.94247 | 1 | F555W | F814W | 413 | 800 |
| j8z007010 | 10.5878 | 41.53735 | 3 | F606W | F814W | 2370 | 2370 |
| J92GB5NFQ | 10.60658 | 41.39225 | 1 | F555W | F814W | 413 | 502 |
| j92gb3dpq | 10.66242 | 41.60167 | 1 | F555W | F814W | 413 | 502 |
| j92gb0cuq | 10.66254 | 41.03394 | 1 | F555W | F814W | 413 | 502 |
| J92GB6ZNQ | 10.6925 | 41.48375 | 1 | F555W | F814W | 413 | 502 |
| j92ga7vlq | 10.73546 | 41.52947 | 1 | F555W | F814W | 151 | 457 |
| J92GB4E9Q | 10.74833 | 41.69314 | 1 | F555W | F814W | 413 | 502 |
| J92GD1FPQ | 10.74846 | 41.12542 | 1 | F555W | F814W | 413 | 502 |
| j92gd4q4q | 10.76183 | 40.99308 | 1 | F555W | F814W | 413 | 502 |
| j92gb8vyq | 10.86433 | 41.66669 | 1 | F555W | F814W | 413 | 502 |
| j8f102cxq | 10.86938 | 41.06218 | 1 | F606W | F814W | 1000 | 1580 |
| J92GD6D5Q | 10.93575 | 41.17469 | 1 | F555W | F814W | 413 | 502 |
| j8mg08071 | 10.94756 | 41.72721 | 3 | F555W | F814W | 825 | 1200 |
| j96q05010 | 10.9882 | 41.52269 | 3 | F606W | F435W | 1840 | 2925 |
| j96q02010 | 11.10045 | 41.4973 | 3 | F606W | F435W | 1860 | 2910 |
| j96q04010 | 11.18791 | 41.47789 | 3 | F606W | F435W | 3315 | 4560 |
| j96q01010 | 11.19553 | 42.07823 | 3 | F606W | F435W | 1850 | 2920 |
| j8o440071 | 11.22637 | 41.52669 | 2 | F606W | F814W | 1830 | 1548 |
| j8mg07071 | 11.28035 | 41.63718 | 3 | F555W | F435W | 825 | 1200 |

# TABLE 2

## MEAN MEASUREMENT ERRORS
## AS A FUNCTION OF MAGNITUDE AND COLOR

| V | Verr | (B – V)err | (V – I)err |
|---|------|------------|------------|
| 15 | 0.01 | 0.01 | 0.32 |
| 16 | 0.03 | 0.05 | 0.03 |
| 17 | 0.03 | 0.04 | 0.18 |
| 18 | 0.04 | 0.07 | 0.28 |
| 19 | 0.06 | 0.06 | 0.24 |
| 20 | 0.11 | 0.12 | 0.22 |
| 21 | 0.12 | 0.14 | 0.22 |
| 22 | 0.23 | 0.39 | 0.23 |
| 23 | 0.21 | 0.32 | 0.23 |

# TABLE 3
## CLUSTERS PREVIOUSLY IDENTIFIED

| Name | RA | Dec | V | Verr | B | Berr | B-V | I | Ierr | V - I |
|------|----|----|---|------|---|------|-----|---|------|-------|
| B008 | 10.12540 | 41.26899 | 17.00 | 0.04 | 17.83 | 0.03 | 0.84 | | | |
| B010 | 10.13076 | 41.23949 | 16.65 | 0.01 | 17.49 | 0.01 | 0.84 | | | |
| c136 | 10.31879 | 40.65087 | 18.96 | 0.04 | 19.21 | 0.05 | 0.25 | | | |
| c137 | 10.32659 | 40.65050 | 19.54 | 0.18 | 19.92 | 0.10 | 0.38 | | | |
| I-34 | 10.42351 | 40.86716 | 19.41 | 0.06 | 20.08 | 0.04 | 0.67 | | | |
| B031D | 10.43581 | 40.85639 | 18.46 | 0.04 | 18.92 | 0.04 | 0.46 | | | |
| B049 | 10.43972 | 40.83202 | 17.83 | 0.03 | 18.37 | 0.03 | 0.54 | | | |
| I-36 | 10.44723 | 40.85249 | 20.16 | 0.07 | 21.20 | 0.08 | 1.04 | | | |
| B253 | 10.45694 | 40.88346 | 19.42 | 0.03 | 19.98 | 0.04 | 0.56 | | | |
| B057 | 10.46986 | 40.86823 | 17.65 | 0.02 | 18.40 | 0.02 | 0.74 | | | |
| B069 | 10.52268 | 41.43591 | 18.56 | 0.05 | | | | 17.91 | 0.09 | 0.65 |
| BO82 | 10.56578 | 41.02058 | 15.54 | 0.01 | | | | 13.65 | 0.01 | 1.89 |
| c200 | 10.57332 | 40.92097 | 19.22 | 0.09 | | | | 18.30 | 0.07 | 0.92 |
| B088 | 10.58760 | 41.53736 | 15.47 | 0.01 | | | | 14.01 | 0.01 | 1.46 |
| B091 | 10.59064 | 41.36819 | 18.09 | 0.04 | | | | 17.42 | 0.05 | 0.67 |
| B093 | 10.59661 | 41.36217 | 16.82 | 0.05 | | | | 15.63 | 0.05 | 1.20 |
| B094 | 10.60413 | 40.95485 | 15.59 | 0.01 | | | | 14.39 | 0.01 | 1.19 |
| B056D | 10.61819 | 41.57427 | 18.92 | 0.03 | | | | 17.53 | 0.04 | 1.39 |
| B067D | 10.66196 | 41.61187 | 19.29 | 0.06 | | | | 18.35 | 0.10 | 0.94 |

| Name | RA | Dec | V | Verr | B | Berr | B-V | I | Ierr | V-I |
|---|---|---|---|---|---|---|---|---|---|---|
| c232 | 10.66636 | 41.03948 | 20.86 | 0.35 | | | | 19.92 | 0.29 | 0.95 |
| B130 | 10.70352 | 41.49801 | 17.05 | 0.03 | | | | 15.66 | 0.02 | 1.39 |
| B137 | 10.72515 | 41.53742 | 17.99 | 0.04 | | | | 16.54 | 0.04 | 1.45 |
| II-113 | 10.72948 | 40.99620 | 19.23 | 0.06 | | | | 18.45 | 0.10 | 0.78 |
| B140 | 10.74456 | 41.14800 | 18.06 | 0.04 | | | | 16.82 | 0.04 | 1.24 |
| B087D | 10.74525 | 41.15242 | 17.64 | 0.02 | | | | 16.59 | 0.03 | 1.05 |
| B141 | 10.74725 | 41.54663 | 17.03 | 0.02 | | | | 15.81 | 0.02 | 1.22 |
| B091D | 10.75618 | 41.50491 | 15.44 | 0.02 | | | | 14.27 | 0.02 | 1.17 |
| B266 | 10.76432 | 41.67529 | 18.48 | 0.03 | | | | 17.00 | 0.01 | 1.48 |
| B174 | 10.87643 | 41.64898 | 15.52 | 0.01 | | | | 14.26 | 0.01 | 1.26 |
| I-66 | 10.85047 | 41.64513 | 20.37 | 0.13 | | | | 19.04 | 0.12 | 1.33 |
| c259 | 10.89531 | 41.16478 | 16.48 | 0.06 | | | | 16.05 | 0.09 | 0.43 |
| B198 | 10.95866 | 41.53125 | 18.01 | 0.05 | 18.89 | 0.09 | 0.88 | | | |
| I-74 | 10.95903 | 41.69659 | 19.50 | 0.04 | | | | 18.45 | 0.06 | 1.05 |
| B201 | 10.97008 | 41.16607 | 16.24 | 0.03 | | | | 15.20 | 0.03 | 1.04 |
| B203 | 10.98252 | 41.54302 | 16.79 | 0.02 | 17.68 | 0.02 | 0.89 | | | |
| B206 | 10.99418 | 41.50498 | 15.02 | 0.01 | 15.85 | 0.01 | 0.83 | | | |
| B213 | 11.01455 | 41.51070 | 16.89 | 0.00 | 17.84 | 0.06 | 0.95 | | | |
| B215 | 11.02659 | 41.52876 | 17.24 | 0.02 | 18.22 | 0.02 | 0.97 | | | |
| B220 | 11.08072 | 41.50987 | 16.54 | 0.01 | 17.31 | 0.02 | 0.77 | | | |
| BO224 | 11.11256 | 41.48057 | 15.23 | 0.01 | 15.99 | 0.01 | 0.76 | | | |
| M047 | 11.15783 | 41.48106 | 19.42 | 0.05 | 20.30 | 0.05 | 0.89 | | | |
| B231 | 11.16090 | 41.46298 | 17.28 | 0.02 | 18.13 | 0.03 | 0.85 | | | |
| B234 | 11.19340 | 41.48818 | 16.85 | 0.02 | 17.76 | 0.03 | 0.90 | | | |
| B366 | 11.19428 | 42.06411 | 16.27 | 0.03 | 17.04 | 0.04 | 0.77 | | | |
| B367 | 11.19632 | 42.09233 | 18.13 | 0.03 | 18.56 | 0.02 | 0.43 | | | |
| B255D | 11.20214 | 42.10382 | 18.25 | 0.07 | 19.19 | 0.06 | 0.95 | | | |
| c406 | 11.21377 | 41.49024 | 20.30 | 0.04 | 20.37 | 0.03 | 0.03 | | | |
| III-159 | 11.22247 | 41.53220 | 20.40 | 0.08 | | | | 19.39 | 0.10 | 1.01 |
| c305 | 11.25343 | 41.51674 | 18.98 | 0.18 | | | | 18.51 | 0.08 | 0.47 |
| c311 | 11.29325 | 41.61290 | 18.32 | 0.06 | 18.23 | 0.05 | -0.09 | | | |
| c312 | 11.29936 | 41.62017 | 18.14 | 0.10 | 0.08 | 0.01 | 0.28 | | | |
| c313 | 11.30592 | 41.62668 | 17.58 | 0.02 | 18.05 | 0.02 | 0.48 | | | |

NOTE: The cluster names are from the RBC (Galleti et al 2004), except for names prefixed by I, II, and III which are from the three tables in Kim et al (2007) and those prefixed by c, which are from Hodge (1979)

TABLE 4

NEW STAR CLUSTERS

| Name | RA | Dec | V | Verr | B | Berr | B-V | I | Ierr | V-I |
|---|---|---|---|---|---|---|---|---|---|---|

| | | | | | | | | | |
|---|---|---|---|---|---|---|---|---|---|
| KHM31-284 | 9.89001 | 40.77265 | 21.75 | 0.22 | | | | 20.16 | 0.31 | 1.59 |
| KHM31-285 | 9.89011 | 40.77258 | 21.76 | 0.16 | | | | 20.19 | 0.16 | 1.57 |
| KHM31-286 | 9.89566 | 40.81459 | 22.28 | 0.29 | | | | 21.66 | 0.21 | 0.62 |
| KHM31-287 | 9.92680 | 40.77808 | 22.77 | 0.14 | | | | 20.66 | 0.11 | 2.11 |
| KHM31-288 | 10.10328 | 41.24350 | 23.16 | 0.27 | 23.87 | 0.27 | 0.70 | | | |
| KHM31-289 | 10.10461 | 41.25509 | 21.87 | 0.19 | 23.02 | 0.20 | 1.15 | | | |
| KHM31-290 | 10.13054 | 41.22922 | 21.12 | 0.10 | 21.93 | 0.11 | 0.81 | | | |
| KHM31-291 | 10.16101 | 41.26554 | 20.43 | 0.08 | 21.25 | 0.08 | 0.82 | | | |
| KHM31-292 | 10.29633 | 40.64710 | 21.44 | 0.09 | 22.17 | 0.11 | 0.73 | | | |
| KHM31-293 | 10.29917 | 40.63040 | 21.07 | 0.15 | 21.50 | 0.12 | 0.43 | | | |

NOTE. – Table 4 is published in its entirety in the electronic edition of the *PASP*. A portion is shown here for guidance regarding its form and content.